\documentclass[preprint]{IEEEtran}

\usepackage{amsmath}
\usepackage{cite}
\usepackage{graphicx}% Include figure files
\usepackage{caption}
\usepackage{subcaption}
\usepackage{dcolumn}% Align table columns on decimal point
\usepackage{bm}% bold math
\usepackage[mathlines]{lineno}% Enable numbering of text and display math
\usepackage{xcolor}
\usepackage{hyperref}

\def\CC{{C\nolinebreak[4]\hspace{-.05em}\raisebox{.4ex}{\tiny\bf ++}}}

\begin{document}

\title{Proton light yield of fast plastic scintillators for neutron imaging}

\author{
J.~J.~Manfredi, B.~L.~Goldblum, T.~A.~Laplace, G.~Gabella, J.~Gordon, A.~O'Brien, S.~Chowdhury, J.~A.~Brown, E.~Brubaker,~\IEEEmembership{Member,~IEEE,}%

\thanks{This work was performed under the auspices of the U.S. Department of Energy National Nuclear Security Administration by Lawrence Berkeley National Laboratory under Contract DE-AC02-05CH11231 and through the Nuclear Science and Security Consortium under Award No. DE-NA0003180. Sandia National Laboratories is a multimission laboratory managed and operated by National Technology and Engineering Solutions of Sandia LLC, a wholly owned subsidiary of Honeywell International Inc., for the U.S. Department of Energy's National Nuclear Security Administration under contract DE-NA0003525.}
\thanks{J.J. Manfredi, B.L. Goldblum, T.A. Laplace, G. Gabella, J. Gordon, A. O'Brien, S. Chowdhury are with the Department of Nuclear Engineering at the University of California, Berkeley.}
\thanks{J.A. Brown was with Sandia National Laboratories, Livermore, California, and is now with the Department of Nuclear Engineering at the University of California, Berkeley.}
\thanks{E. Brubaker is with Sandia National Laboratories, Livermore, California}
}

\maketitle

\begin{abstract}
Plastic organic scintillators have been tailored in composition to achieve ultra-fast temporal response, thereby enabling the design and development of fast neutron detection systems with high timing resolution. Eljen Technology's plastic organic scintillators---EJ-230, EJ-232, and EJ-232Q---are prospective candidates for use in emerging neutron imaging systems, where fast timing is paramount. To support the neutron response characterization of these materials, the relative proton light yields of EJ-230, EJ-232, and EJ-232Q were measured at the 88-Inch Cyclotron at Lawrence Berkeley National Laboratory. Using a broad-spectrum neutron source and a double time-of-flight technique, the proton light yield relations were obtained over a proton recoil energy range of approximately 300~keV to 4~MeV. The EJ-230, EJ-232, and EJ-232Q scintillators exhibited similar proton light yield relations to each other as well as to other plastic scintillators with the same polymer base material. A comparison of the relative proton light yield of different sized cylindrical EJ-232 and EJ-232Q scintillators also revealed consistent results. This work provides key input data for the realistic computational modeling of neutron detection technologies employing these materials, thereby supporting new capabilities in near-field radionuclide detection for national security applications. 
\end{abstract}


\section{Introduction}

Plastic organic scintillators are durable with rapid rise and decay times, making them useful in a wide variety of applications in nuclear and particle physics, national security, and medical imaging \cite{GatuJohnson, Zhao, Green, Uchiyama, Moskal}. For fast neutron detection using these materials, the dominant mechanism for light generation is n-p elastic scattering, where energy is transferred to the scintillating medium via the recoiling proton. The proton light yield (PLY), or light emitted as a function of proton recoil energy, is a critical input in the determination of detection efficiencies for absolute calibration of neutron time-of-flight (TOF) spectra and for reconstruction in kinematic neutron imaging algorithms \cite{Mirfayzi, Sweany}. This work provides a measurement of the PLY relation for the EJ-230, EJ-232, and EJ-232Q fast plastic organic scintillators developed by Eljen Technologies. 

These scintillators are composed of a polyvinyltoluene (PVT) solvent with one or more proprietary fluors, which give rise to the unique properties of each material listed in Table~\ref{tab:scintprops}. For the EJ-232Q sample used in this work, 0.5\% benzophenone was added to the medium as a quenching agent, resulting in an ultra-fast rise time and a sub-ns prompt decay constant. The temporal properties of these materials make them particularly attractive for use in prospective single-volume neutron scatter camera systems, where two n-p elastic scattering events are detected within a contiguous volume \cite{braverman, weinfurther}. In such an approach, the energy and angle of the incident neutron is determined by measuring the position, time, and energy deposition associated with the double-scatter event. Specifically, the energy of the incident neutron, $E_n$, is determined via conservation of energy: $E_n = E_n^{\prime}+E_p$. The scattered neutron energy, $E_n^{\prime}$, is inferred from the position of and time-of-flight between interaction points in the contiguous volume, necessitating fast scintillators and excellent timing resolution. The proton recoil energy, $E_p$, is obtained by converting the measured scintillation light into proton energy via the empirically-determined PLY relation. The scattering angle of the incident neutron is then determined from $E_p$ and $E_n^{\prime}$ via n-p elastic scattering kinematics. The PLY measurements obtained in this work facilitate the design and operation of single volume neutron scatter camera systems relying on EJ-230, EJ-232, or EJ-232Q as the detection medium. Indeed, recent work from Sweany et al.\ suggests EJ-230 as a top candidate for use in the construction of an optically segmented single-volume neutron scatter camera, based on both position and timing resolution performance in test scintillator bars with dual-ended readout \cite{Sweany}. 

\begin{table*}
\centering
\caption{Scintillator efficiencies, rise times, and decay times as reported by Eljen Technologies \cite{230spec,232spec}.}

\label{tab:scintprops}
\renewcommand{\arraystretch}{1.2}
\begin{tabular}{ccccc}
\hline
\hline
Material & Scintillation Efficiency & Rise Time  & Decay Time  & Commercial \\
 & (photons/MeVee)  & (ps) & (ns) & Equivalents \\
\hline
EJ-230 & 9700 & 500 &1.5 & BC-420, Pilot U2 \\
EJ-232 & 8400 & 350 & 1.6 & BC-422, NE-111A \\
EJ-232Q$^{\ast}$ & 2900 & 110 & 0.7 & BC-422Q \\
\hline
\hline
\end{tabular}
\vspace{1ex}

 {\raggedright \hspace{24ex} $\ast$ 0.5\% benzophenone.  \par}
\end{table*}

This work also addresses a recent controversial claim in the literature that the relative PLY of a scintillator depends upon the size of the scintillating volume. Enqvist et al.\ measured the relative PLY of three EJ-309 organic scintillator cells of different sizes and asserted that the relative PLY of EJ-309 varied with scintillator size due to self-attenuation of the scintillation light \cite{enqvist}. The relative PLY provides the light output for proton recoils relative to that for electron recoils and the attenuation of the scintillation light as it travels through a medium is the same regardless of the means by which the light was produced. Excluding edge effects, the impact of a change in scintillator size should not manifest as a change in the relative PLY, and rather should be accounted for in the light unit calibration. A size-dependent relative PLY would have significant consequences in any application of organic scintillators for neutron detection, as such an effect would require the PLY to be independently evaluated for a particular scintillator geometry. To investigate this issue, the relative proton light yield of two different sized cylindrical EJ-232 and EJ-232Q scintillators was measured. 

The experimental methods and setup are detailed in Sec.~\ref{exp}. Section~\ref{analysis} describes the data analysis and uncertainty quantification procedures. In Sec.~\ref{results}, the measured PLY relations of EJ-230, EJ-232, and EJ-232Q are reported and compared with literature measurements where available. The EJ-232 and EJ-232Q PLY relations for two different sized scintillators are reported and the implications for applications are addressed. Concluding remarks are provided in Sec.~\ref{conc}.

\section{Experimental Methods}
\label{exp}

The double-time-of-flight (dTOF) method of Brown et al.\ was recently developed for characterizing the relative PLY of organic scintillators \cite{Brown18,BrownThesis}. In accordance with this approach, the scintillator of interest, i.e., the {\it target scintillator}, was placed in an intense neutron beam generated at the 88-Inch Cyclotron at Lawrence Berkeley National Laboratory. The cyclotron accelerated a 16 MeV deuteron beam onto a 3-mm-thick beryllium target, resulting in a high-flux and broad-spectrum neutron source \cite{Harrig18}.

When neutrons elastically scatter off of protons in the organic scintillator, the recoiling protons excite the scintillation medium. The goal of this work is to associate the recoil energy of the proton with the light output in the target scintillator (i.e., the PLY). To accomplish this, the energy and angle of the scattered neutron are used to kinematically determine the recoiling proton energy. For this purpose, an array of eleven EJ-309 liquid organic scintillators, i.e., {\it scatter scintillators}, surrounded the target at fixed positions out-of-beam during the experiment. Using the pulse-shape discrimination (PSD) properties of the EJ-309 scintillators, neutron interactions were identified in the scatter scintillators on an event-by-event basis. 

The incoming neutron energy for each event was determined using the time-of-flight (TOF) between the cyclotron RF signal and the target scintillator. Similarly, the outgoing scattered neutron energy was calculated using the TOF between the target scintillator and the corresponding scatter scintillator. The proton recoil energy $E_p$ was then deduced from the scattered-neutron energy $E_n'$ and the known scattered-neutron angle relative to the neutron beam axis, $\theta$, via n-p elastic scattering kinematics: 
\begin{equation}
\label{eq:out}
E_p = E_n' \tan^{2}(\theta).
\end{equation}
By associating this recoil energy with the corresponding light response in the target scintillator for each event, a continuous measurement of the target scintillator PLY was obtained. 

\begin{figure}
	\center
	\includegraphics[width=0.5\textwidth]{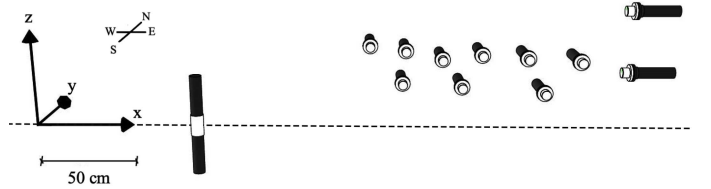}
	\caption{Schematic of the detector array configuration for the EJ-230 proton light yield measurement. The neutron beam traveled along the $x$-axis through the target scintillator with dual-ended readout, while scatter detectors were positioned at forward angles to detect elastically scattered neutrons.} 
	\label{fig:schematic}
\end{figure}

Figure \ref{fig:schematic} shows a diagram of the EJ-230 experimental setup, which is similar in layout to those employed in the EJ-232 and EJ-232Q experiments. The $x$-coordinate corresponded to the distance from the west wall of the experimental cave, the $y$-coordinate to the distance from beamline-center in the north direction, and the $z$-coordinate to the vertical distance from beamline-center. The detector positions in each experiment are detailed in Appendix~\ref{app-exp}. The target scintillators were right cylinders (5.08~cm dia.) with each end of each cylinder optically coupled to a photomultiplier tube (PMT) using BC-630 silicone grease. The EJ-230 scintillator was 5.08~cm in height, and cylinders of two different heights (2.54~cm and 5.08~cm) were measured for both the EJ-232 and EJ-232Q media. Each target scintillator was wrapped in at least 10 layers of Teflon tape to maximize the reflectivity of the detector housing \cite{Janacek2012}. The scatter scintillators (5.08~cm dia. x 5.08~cm h.) were composed of right-cylindrical cells of thin aluminum housing filled with EJ-309 and coupled to a single PMT via a borosilicate glass window and BC-630 silicone grease. All PMTs used in these measurements were obtained from Hamamatsu Photonics (either Type No. 1949-50 or 1949-51), and were negatively biased using either a CAEN R1470ET or CAEN NDT1470 power supply. Further details on the target scintillator PMTs are provided in Table~\ref{pmtTable} in Appendix~\ref{app-exp}.

The PMT signals were connected to a CAEN V1730 500~MS/s digitizer. Full waveforms were recorded within a 670-ns acquisition window for all scintillator signals as well as for the cyclotron RF control signal. The data acquisition was triggered on a coincidence between both of the target scintillator PMT signals as well as at least one of the scatter scintillator signals within a 400~ns coincidence window. The CAEN digital constant fraction discrimination algorithm, with a 75\% fraction and a 4~ns delay, was used to determine the arrival time of the scintillator signals. The timing pick-off for the cyclotron RF signal was accomplished using leading edge discrimination. All waveforms were written to disk (using the ROOT file format \cite{Brun97}) in list mode with global time stamps to enable event reconstruction.%V2

\section{Data Analysis}
\label{analysis}

The raw PMT waveforms were processed using a custom \CC\ ROOT-based analysis framework. Integration of each individual waveform provided a measure of the light collected in a particular event. The integration lengths were chosen to collect at least 95\% of the scintillation light in the 670 ns acquisition window. For the EJ-232Q scintillator, the integration length was 140~ns, while for both the EJ-230 and EJ-232 scintillators, the integration length was 180~ns. As in previous studies, a 300~ns integration length was used for the EJ-309 scatter cells \cite{Laplace18, Brown18}. For the target scintillators, the pulse integrals for each of the target PMTs were combined using a geometric average to obtain an event-by-event light output measurement independent of interaction position \cite{Knoll2010}. The linearity of the response of each target PMT was evaluated using a specialized pulsed-LED circuit based on the approach of Friend et al. (see Appendix \ref{app-exp} for details) \cite{Friend11, BrownThesis}. 

The light-output unit for the PLY relations presented in this work is defined relative to the light produced by a 477~keV recoil electron. Electron-equivalent light units, although common in the literature, implicitly assume electron light proportionality. Several recent studies have shown that this assumption is not valid for plastic organic scintillators in the energy range relevant for this measurement \cite{Nassalski08, Payne11}. The use of a light-output unit relative to the light produced by a 477~keV recoil electron ensures that the results presented here capture the behavior of the PLY without bias introduced by potential electron light yield nonlinearity.%V1

\begin{figure}
	\center
	\includegraphics[width=0.5\textwidth]{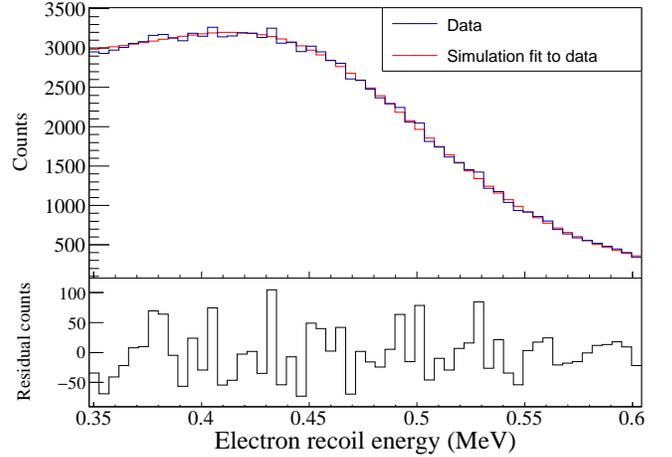}
	\caption{(Color online) The top panel shows the results from a pulse-integral calibration using a $^{137}$Cs source incident on the EJ-232Q (5.08~cm h.) scintillator. The experimental spectrum (blue) and a simulated spectrum fit to the data and folded with a PMT response (red) are shown. The shape of the data are well reproduced by the fit, as evidenced by the residual plot on the bottom panel.}
	\label{fig:pulseint}
\end{figure}

Light-output calibrations for each target scintillator were performed with a $^{137}$Cs source using the Compton edge (477~keV) of the 662~keV $\gamma$-ray. This source was chosen for its strong mono-energetic line that lay within the dynamic range of this experiment. The Cs source was positioned at a distance of at least 5 cm from the target scintillator for all but the 2.54 cm EJ-232 sample, where it was taped directly to the face of the cylinder during calibration. The resulting pulse-integral spectra were fit using $\chi^2$-minimization to simulated deposited-energy spectra (convolved with a PMT-resolution function \cite{dietze}) to determine the channel number corresponding to the Compton edge. An example light output calibration spectrum is shown for the 5.08~cm EJ-232Q scintillator in Fig.~\ref{fig:pulseint}. The $\chi^2$-minimization was implemented using the SIMPLEX and MIGRAD algorithms from the ROOT MINUIT2 package. In both the light-output calibrations and the neutron-beam data, edge effects due to escaping recoil particles are negligible due to the small size of the recoil-particle ranges (approximately 350 $\mu$m for a 5 MeV proton) compared to the scintillator dimensions.%V2

To obtain the proton recoil energy (as described in Eq.~\ref{eq:out}), timing calibrations were performed for both the incoming and outgoing neutron TOF. The incoming neutron TOF, $t_{\text{inc}}$, was calculated as follows:
\begin{equation}
\label{eq:inccalc}
t_{\text{inc}} = \frac{t_1 + t_2}{2} - t_{\text{RF}} - t_{\text{cal}} + (n \times T_{\text{cyc}}),
\end{equation}
where $t_1$ and $t_2$ are the signal arrival times in each of the two target scintillator PMTs, $t_{\text{RF}}$ is the time of the cyclotron RF signal, $t_{\text{cal}}$ is a time-calibration constant, ${T_\text{cyc}}$ is the cyclotron pulse period (158.5 ns), and $n$ is an integer to account for the possibility that the incoming neutron was generated by different cyclotron pulses. The determination of $n$ will be discussed later in this section. The determination of $t_{\text{cal}}$ was accomplished using the photon flash produced by deuterons impinging on the beryllium target and the known flight-path between the beryllium target and the target scintillator. The signal from the prompt-$\gamma$ flash was fit with a Gaussian distribution plus a linear background term. Given the speed of light and the known flight-path distance, the centroid of the fit was used to determine $t_{\text{cal}}$. The width of the prompt-$\gamma$ distribution varies between 2.5 and 7~ns depending on the quality of the beam tune and is a significant contributor to the incoming-neutron energy uncertainty.%V1

Similarly, the outgoing neutron TOF was calibrated by examining time differences between $\gamma$-ray events in the target scintillator and each individual scatter cell. For scatter scintillator $i$, the outgoing neutron TOF, $t_{i,\text{out}}$, is given by:
\begin{equation}
\label{eq:outcalc}
t_{i,\text{out}} = t_{i} - \frac{t_1 + t_2}{2} - t_{i,\text{cal}},
\end{equation}
where $t_{i}$ is the measured time of the signal from the $i$th scatter scintillator and $t_{i,\text{cal}}$ is the corresponding time-calibration constant. After gating on $\gamma-\gamma$ coincidences using PSD in each EJ-309 scatter cell, histograms were made of the coincident TOF for each target-scatter pair. Each time-calibration constant $t_{i,\text{cal}}$ was determined from the centroid of the prominent peak in the corresponding histogram. As in the case of the incoming TOF, the width of the peak (on average approximately 400~ps) contributes to the scattered-neutron energy uncertainty.%V2

As this setup overconstrains n-p elastic scattering kinematics in the target scintillator, the integer ambiguity, $n$, in Eq.~\ref{eq:inccalc} can be resolved. Using the outgoing neutron TOF and the scattering angle, the expected incoming neutron TOF was calculated on an event-by-event basis and compared to the measured incoming neutron TOF for different values of $n$. All events in which the expected incoming TOF differed from the measured incoming TOF by more than 5\% of the cyclotron period (7.925 ns) were discarded. This additional constraint removes contributions from multiple neutron interactions in the target scintillator, which represent a large potential contributor to systematic bias in edge characterization approaches to PLY measurement \cite{tomanin}.%V1

\begin{figure}
	\center
	\includegraphics[width=0.5\textwidth]{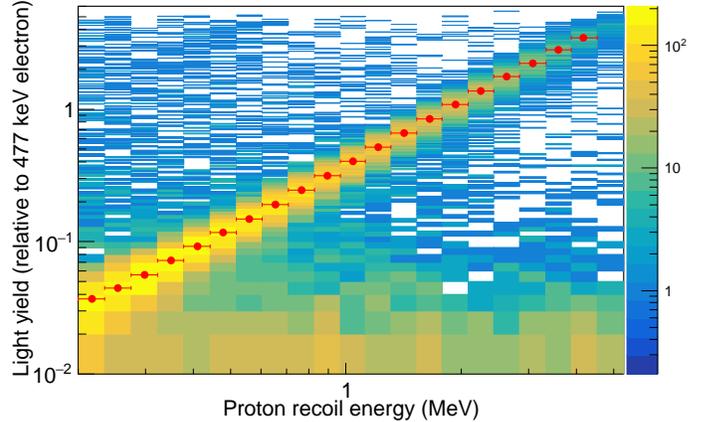}
	\caption{(Color online) 2D histogram of relative light output versus proton recoil energy for the EJ-230 measurement. The binning on the abscissa was set using the proton recoil-energy resolution. The red data points indicate the centroids from fitting projected pulse integral spectra in each bin. The $y$-error bars, which are smaller than the data points, represent the statistical and systematic relative light output uncertainty.}\label{fig:ly2d}
\end{figure}

\begin{figure*}[t!]
	\centering
	\begin{subfigure}{0.5\textwidth}
		\centering
		\includegraphics[width=0.97\textwidth]{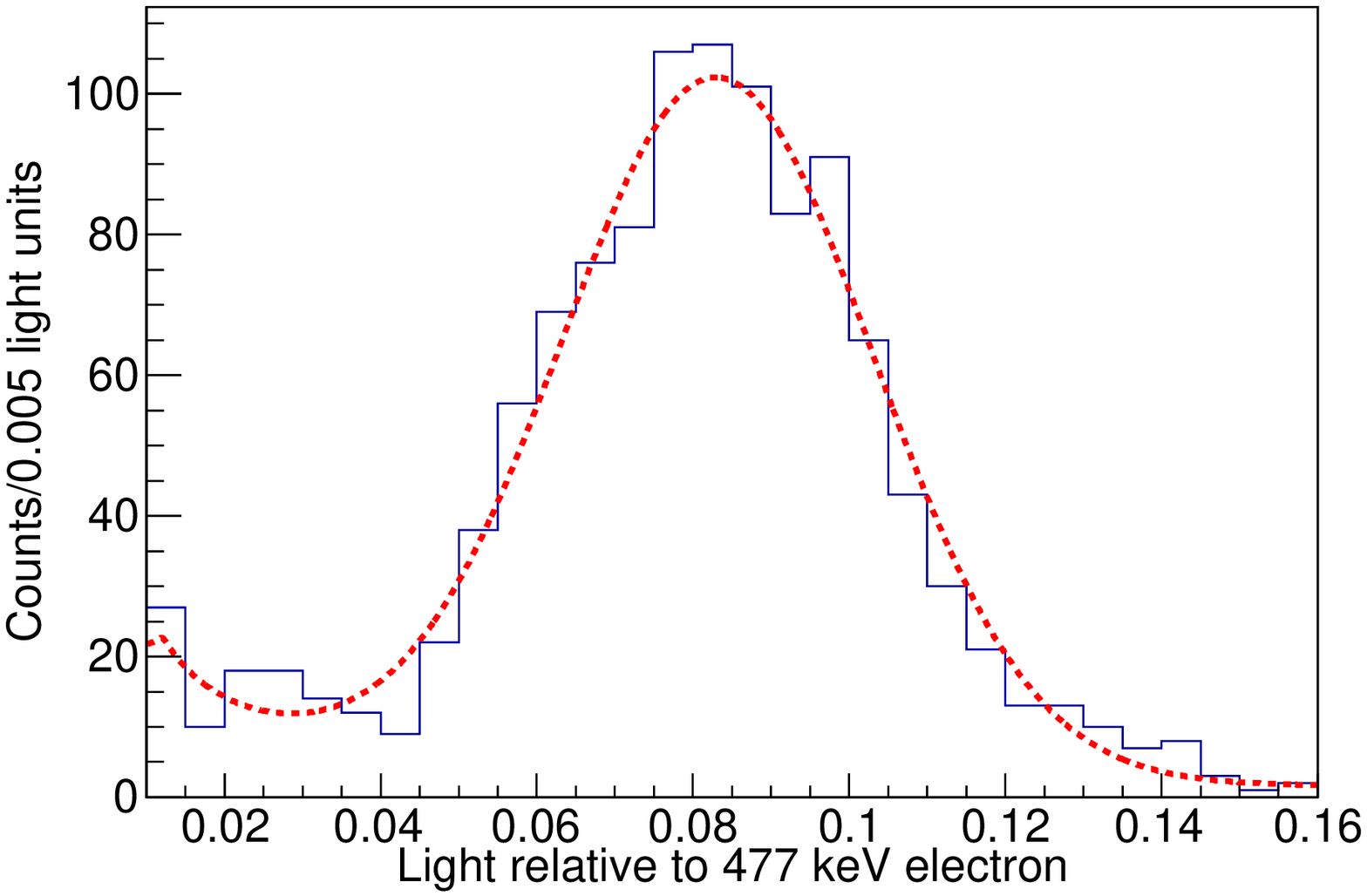}
		\caption{E$_p$ = 412 keV}
	\end{subfigure}%
	~ 
	\begin{subfigure}{0.5\textwidth}
		\centering
		\includegraphics[width=0.97\textwidth]{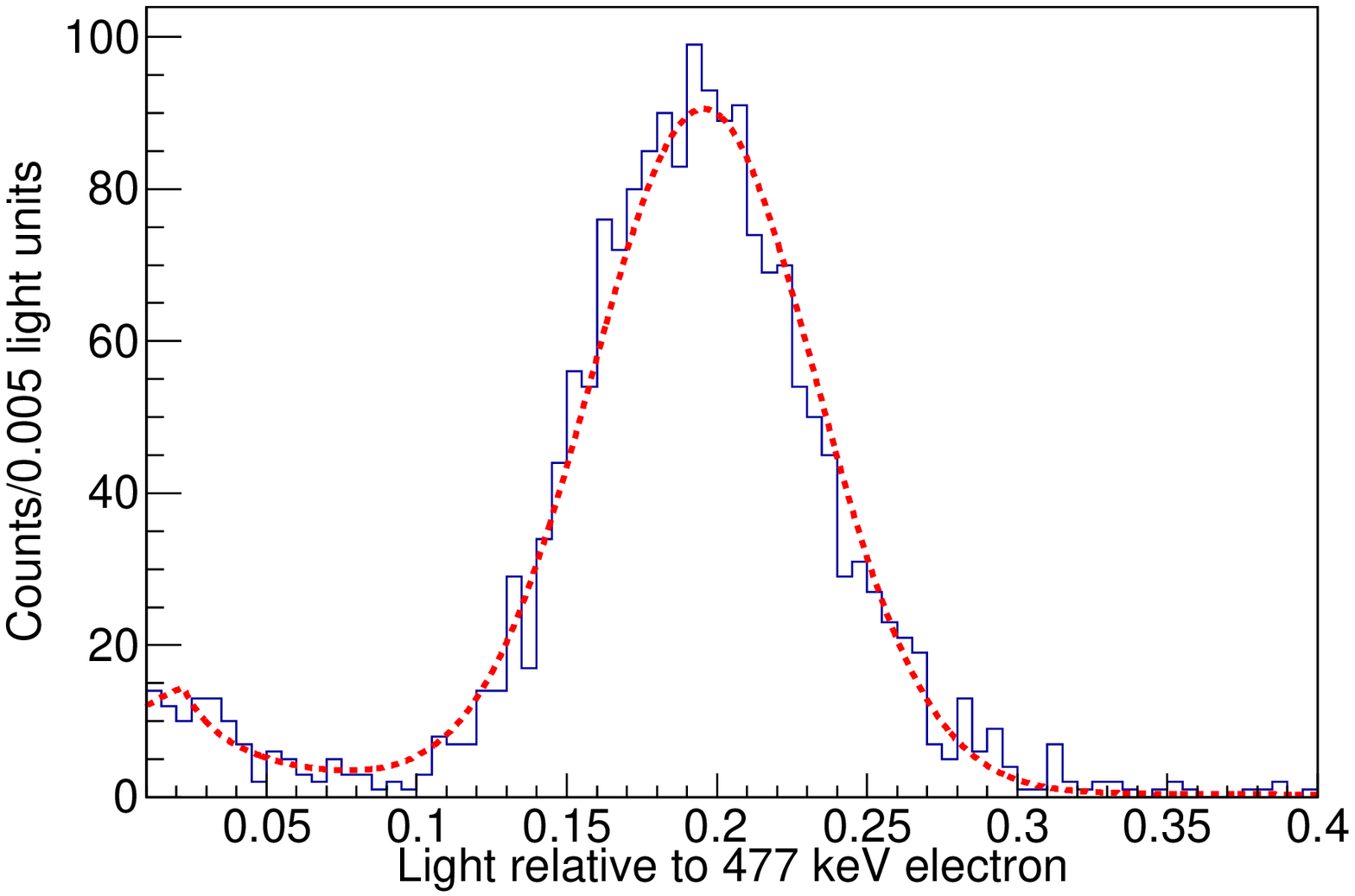}
		\caption{E$_p$ = 718 keV}
	\end{subfigure}
	 
	\begin{subfigure}{0.5\textwidth}
		\centering
		\includegraphics[width=0.97\textwidth]{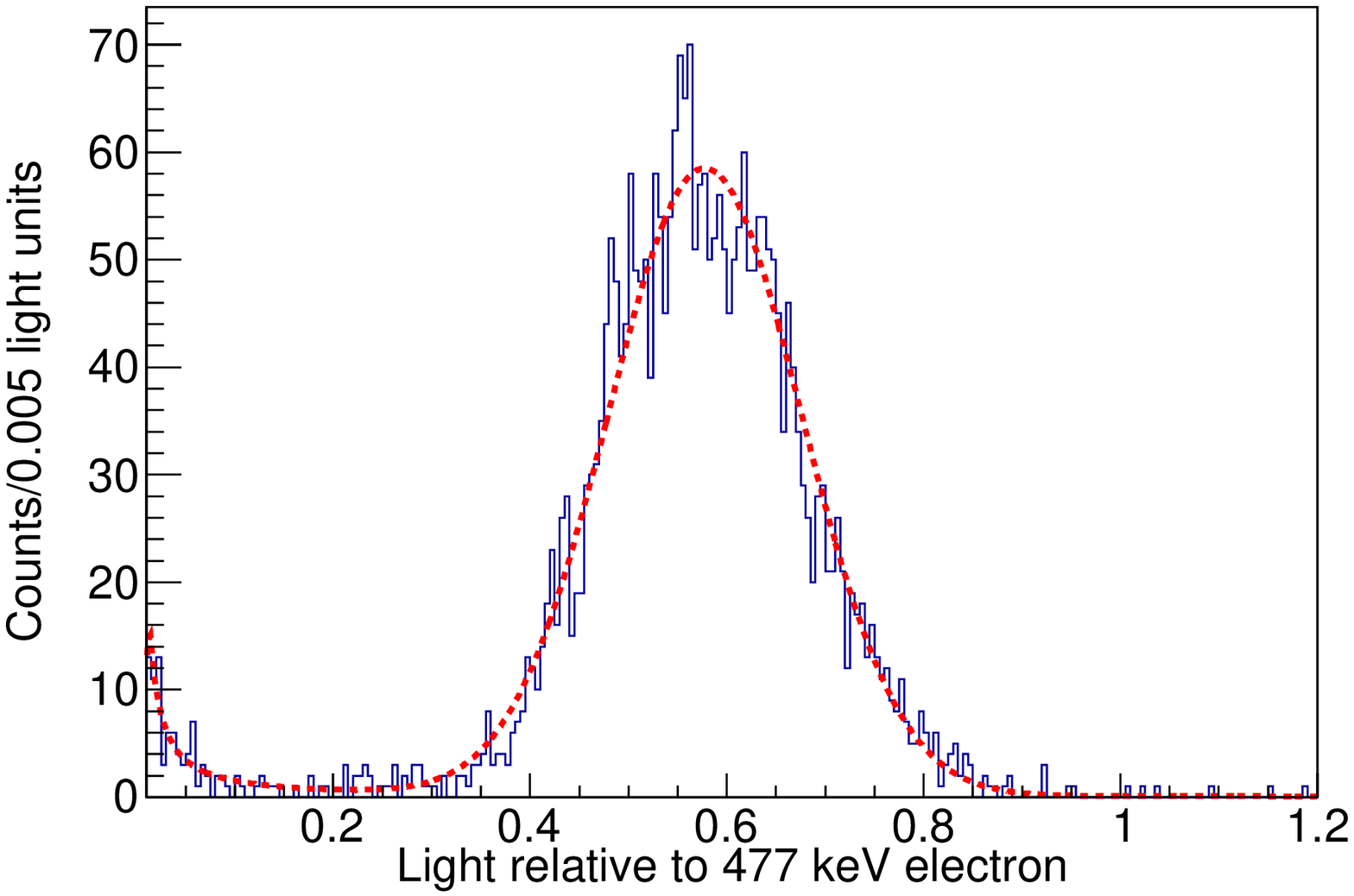}
		\caption{E$_p$ = 1389 keV}
	\end{subfigure}%
	~ 
    \begin{subfigure}{0.5\textwidth}
	   \centering
	   \includegraphics[width=0.97\textwidth]{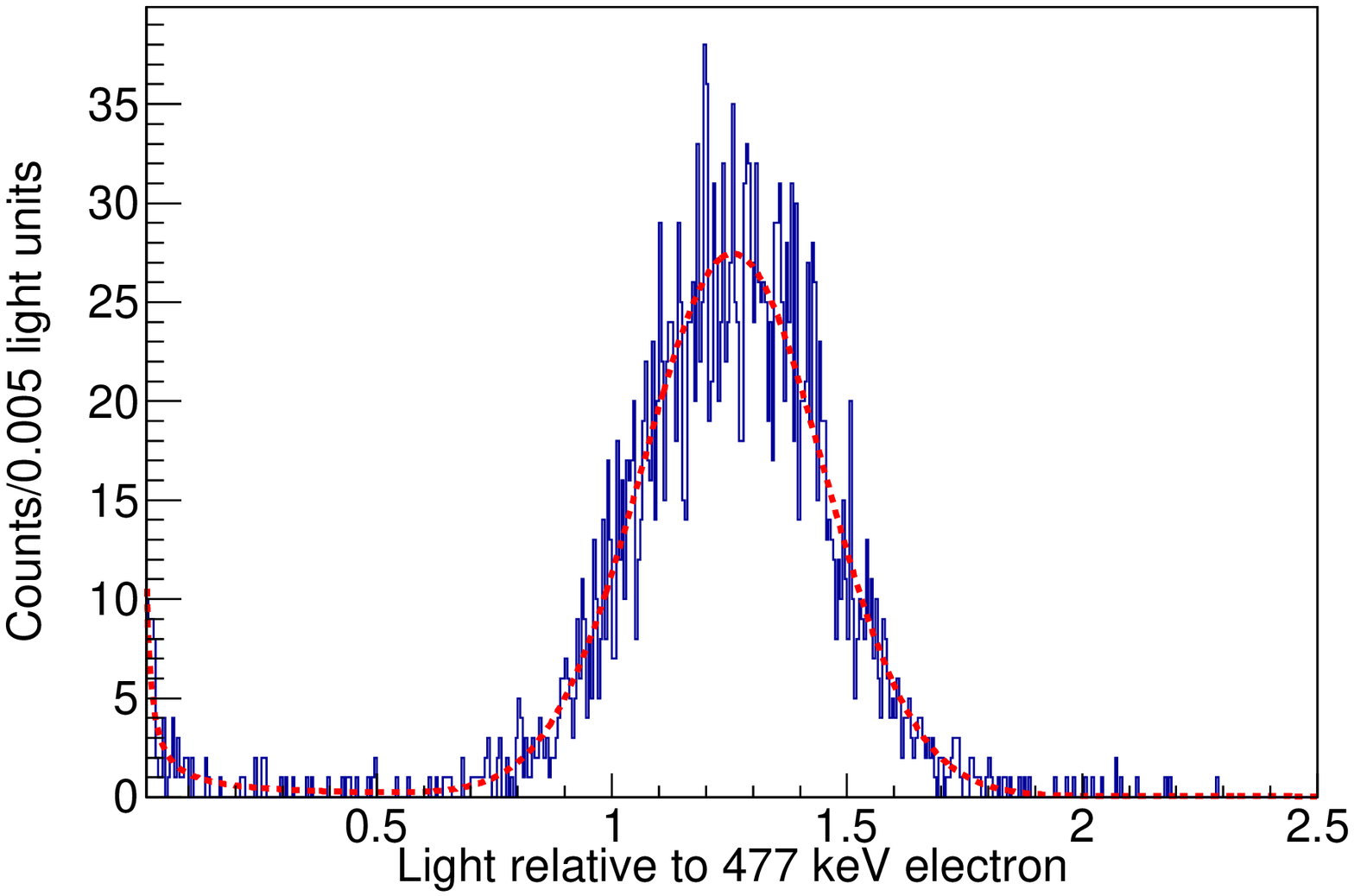}
	   \caption{E$_p$ = 2236 keV}
    \end{subfigure}
	\caption{(Color online) Relative light yield spectra (blue curves) for a given proton-energy bin fit with a piecewise power law and a Gaussian distribution (red dashed curves) for the EJ-232 (5.08~cm h.) measurement. The centroid of the Gaussian distribution corresponds to the mean light production for n-p elastic scattering events within the bin.}
	\label{fig:slices}
\end{figure*}

The proton recoil energy was determined for each event using Eq.~\ref{eq:out}. Figure~\ref{fig:ly2d} shows a histogram of the relationship between light output and proton recoil energy, with a clearly visible kinematic band. Events in this histogram were gated on the observation of neutrons in the scatter cells using PSD. Also, a maximum neutron energy limit was applied using the known incoming-beam energy and the detector configuration. Uncertainty in the proton recoil energy was calculated from a Geant4\mbox{\cite{geant4}} simulation taking into account the uncertainty in the outgoing neutron TOF, the uncertainty in the flight-path length, and the finite size of the detectors (resulting in an angular variance of approximately 0.8 degrees). For each simulated event, the proton recoil energy was reconstructed with the simulated scattered neutron energy and Eq.~{\ref{eq:out}}. This reconstructed proton recoil energy was then compared to the simulated proton recoil energy to calculate the effects of the aforementioned uncertainties on the energy determination. The resulting energy resolution is heteroskedastic, providing an uncertainty that varies as a function of proton recoil energy. The energy resolution is dominated by the angular spread of the detectors, as opposed to the TOF resolution. The binning for the $x$-axis was determined using the proton recoil energy resolution function. Each individual bin was projected onto the light-yield axis and fit using a binned maximum-likelihood estimation with a Gaussian distribution and a continuous piecewise power law. Figure \ref{fig:slices} shows several examples of these individual energy-bin fits.

The uncertainty in the light output was calculated from the following sources: sensitivity to parameters in the light-yield analysis, the stability of the PMT gain over time, and uncertainty in the Compton-edge determination during the pulse-integral calibration. To determine the sensitivity of the light output to analysis parameters, Monte Carlo calculations were performed in which each parameter was varied by sampling from the probability distribution defined by the associated uncertainty. The data reduction was then repeated with these varied parameters to determine the influence on the extracted light-yield. The varied parameters were the incoming/outgoing neutron TOF calibration constants, the measured scatter-scintillator locations, the measured target-scintillator location, and the flight path from the beryllium target to the origin of the coordinate system. Although these parameters directly affect only the proton recoil energy calculation, the fixed binning structure leads to some events in each Monte Carlo trial migrating into adjacent bins as the parameters are varied. As such, the uncertainty in the proton recoil energy is translated to an uncertainty in the light yield as the event composition of each bin changes. To evaluate the PMT gain stability, subsets of the beam data were analyzed to evaluate the light output as a function of time and a variation of 1\% was observed. To evaluate the uncertainty from the pulse-integral calibration, the systematic uncertainty in the Compton-edge determination was estimated to be 1\% by varying the fitted region around the Compton edge, while the statisical uncertainty was negligibly small. In the case of the 2.54 cm EJ-232 sample for which the Cs source was taped directly to the face of the cylinder, an edge-determination uncertainty of 3\% was obtained. This was informed by comparing two sets of calibration data for the 5.08 cm EJ-232 scintillator (one in which the source was taped to the face of the cylinder, and another in which the source was placed 5 cm away) and observing the shift in the Compton edge. Simulations using Geant4 were also performed to study the effect of nonuniform illumination of the cylindrical scintillator. The resulting uncertainties from these three sources were combined in quadrature to arrive at the final light-output uncertainty for each data point.

\section{Results and Discussion}
\label{results}

For the 5.08 cm h.\ x 5.08 cm dia.\ cylindrical scintillator, the measured relative PLY relations of the EJ-230, EJ-232, and EJ-232Q scintillators are shown in Figure \ref{fig:results}, and the corresponding data are tabulated in Appendix~\ref{LYapp}. The three scintillating media have relative PLYs that are similar in shape and magnitude. The difference between any two of the three materials is within approximately two sigma from zero across the full range. As ionization quenching (i.e., quenching of the primary excitation by a high density of excited and ionized molecules) is a primary process occurring in the solvent or host material for solid organic systems, this consistency is expected given that the three materials all employ the same PVT polymer base \cite{birks}.

\begin{figure}
	\center
	\includegraphics[width=0.5\textwidth]{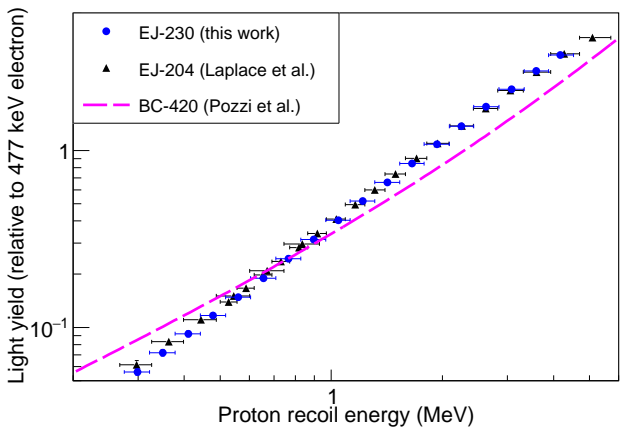}
	\caption{(Color online) Relative proton light yields for EJ-230 from this work (blue circles) and from a measurement of the commercial equivalent BC-420 (magenta dashed line). A recent measurement of EJ-204 \cite{Laplace18}, which has the same polymer solvent as EJ-230, is shown with black triangles. The $x$-error bars represent proton energy bin widths. The $y$-error bars represent the statistical and systematic light output uncertainty. For the EJ-230 and EJ-204 data, some error bars are smaller than the data points. No uncertainties were reported for the BC-420 measurement. \label{fig:pozziComp}}
\end{figure}

Figure~\ref{fig:pozziComp} shows a comparison of the EJ-230 PLY obtained in this work to a measurement by Pozzi et al.\ of the PLY of BC-420 (an EJ-230 commercial equivalent) \cite{Pozzi2004}. Pozzi et al.\ used a $^{252}$Cf source incident on several plastic scintillator detectors and obtained the PLY by quantifying the electron-equivalent light output corresponding to an inferred minimum neutron energy for a given detection threshold \cite{Pozzi2004, Mihalczo}. There is a clear discrepancy between the two measurements, as the present work shows significantly higher relative light yield for several-MeV proton recoil energies and lower relative light yield below 1 MeV proton recoils. The dTOF technique used in this work enables a direct measurement of the relative PLY compared to the method used by Pozzi et al., which infers the proton energy from neutron efficiency curves without fully accounting for systematic uncertainties. The black triangles show a dTOF measurement by Laplace et al.\ of the PLY of EJ-204, a fast plastic organic scintillator from Eljen Technology that also employs a PVT solvent \cite{Laplace18}. The EJ-230 result obtained in this work is consistent with the PLY of the EJ-204 scintillator within the estimated uncertainties.

\begin{figure}
	\center
	\includegraphics[width=0.5\textwidth]{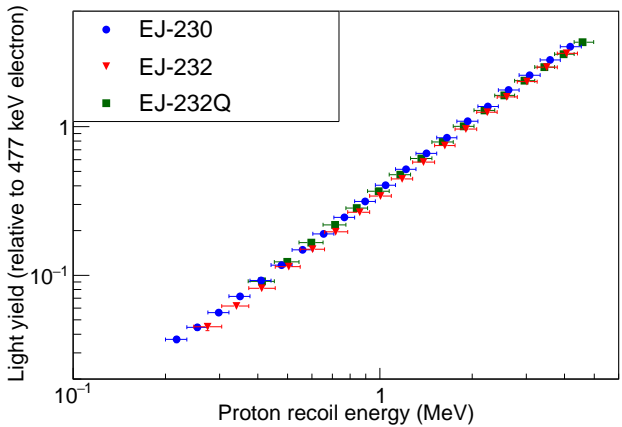}
	\caption{(Color online) Relative proton light yield relations for 5.04 cm h.\ EJ-230 (blue circles), EJ-232 (red triangles), and EJ-232Q (green squares). The $x$-error bars represent proton energy bin widths. The $y$-error bars represent the statistical and systematic light output uncertainty. In some cases, the error bars are smaller than the data points. \label{fig:results}}
\end{figure}

\begin{figure}[h!]
	\center
	\includegraphics[width=0.5\textwidth]{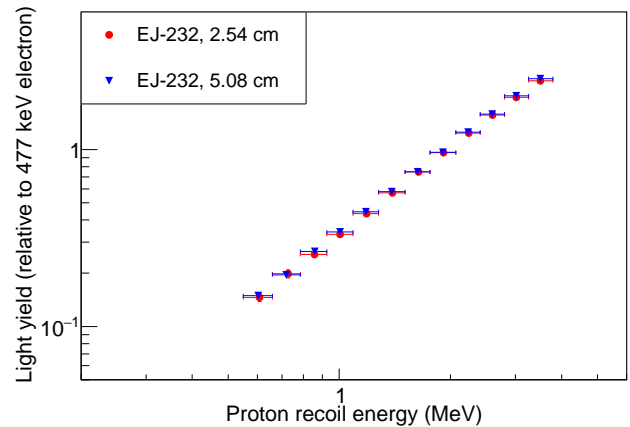}
	\caption{(Color online) The relative proton light yields for the 2.54~cm (red) and 5.08~cm (blue) EJ-232 scintillators. The $x$-error bars represent proton energy bin widths. The $y$-error bars represent the statistical and systematic light output uncertainty. In some cases, the error bars are smaller than the data points. \label{fig:comp_232}}
\end{figure}

\begin{figure}[h!]
	\center
	\includegraphics[width=0.5\textwidth]{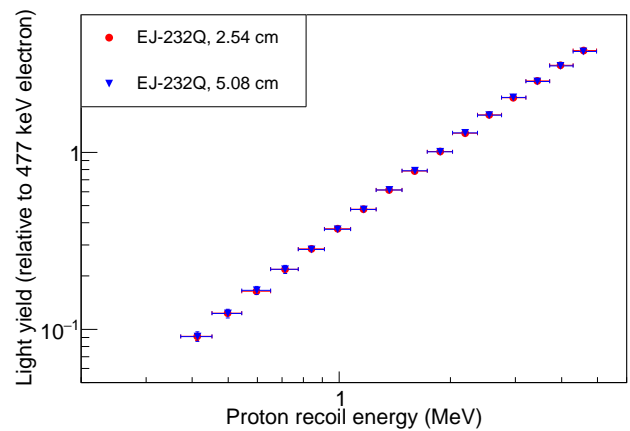}
	\caption{(Color online) The relative proton light yields for the 2.54~cm (red) and 5.08~cm (blue) EJ-232Q scintillators. The $x$-error bars represent proton energy bin widths. The $y$-error bars represent the statistical and systematic light output uncertainty.  In some cases, the error bars are smaller than the data points.\label{fig:comp_232Q}}
\end{figure}

A recent study by Enqvist et al.\ of the EJ-309 organic scintillator suggested that the relative PLY varies depending on the scintillator cell size due to attenuation of the scintillation light \cite{enqvist}. To test this conjecture, PLY measurements of cylindrical EJ-232 and EJ-232Q cells with heights of both 2.54~cm and 5.08~cm were performed. If self-attenuation resulted in size-dependence of the relative PLY (as Enqvist et al.\ argue), then such an effect would be particularly noticeable in EJ-232 and EJ-232Q given the short optical attenuation lengths of the materials ($\sim$17~cm and $\sim$6~cm, respectively) \cite{230spec,Ebran13}. Figures \ref{fig:comp_232} and \ref{fig:comp_232Q} show comparisons of the relative PLY results for the 2.54~cm and 5.08~cm scintillator sizes for EJ-232 and EJ-232Q, respectively. In both cases, the relative PLY measurements for the two cell sizes agree within the estimated uncertainties. The lack of a statistically significant difference between the data obtained from the two different cell sizes presented here is consistent with basic physical intuition of the scintillation process. Attenuation occurs for scintillation light produced by both electron and proton recoils, and therefore should not affect the {\it relative} light yield. Recently, relative PLY measurements obtained with respect to the crystalline axes of multiple-sized stilbene samples  were performed using a coincident kinematic technique \cite{Weldon19}. These measurements also showed no size-dependence.

\section{Summary}
\label{conc}

The relative PLYs of EJ-230, EJ-232, and EJ-232Q were measured using a dTOF technique for proton recoil energies from approximately 300 keV to 4 MeV. The three resulting PLY relations are similar to each other, as well as to recent measurements of EJ-200, EJ-204, and EJ-208 \cite{Laplace18}. This is reasonable considering that ionization quenching is a primary process and these scintillators share the same PVT base material. The EJ-230 result presented here disagrees significantly from the only previous literature measurement (which was of the commercial equivalent BC-420) \cite{Pozzi2004}. Two different cylindrical cell sizes were measured for the EJ-232 and EJ-232Q scintillators, and in both cases the PLY was found to be consistent within uncertainty between the two sizes. The proton light yield relations provided herein support continued development of kinematic neutron-imaging systems, in particular for applications that require excellent timing resolution.

\appendices

\section{Experimental Details}
\label{app-exp}
The detector positions for the measurements of EJ-230, EJ-232, and EJ-232Q are listed in Tables \ref{locationsEJ230}, \ref{locationsEJ232}, and \ref{locationsEJ232Q}, respectively. The origin of the coordinate system is defined at the beamline center along the west wall of the experimental area (see Fig.~\ref{fig:schematic}). The $x$-coordinate was taken as the distance from the west wall, the $y$-coordinate as the distance north from the beamline center, and the $z$-coordinate as the distance above and below the beamline center. The flight path between the beryllium target and the origin of the coordinate system was $646.9~\pm~0.5$~cm along the $x$-axis. The PMT model numbers and bias voltages used for the target scintillators are detailed in Table~\ref{pmtTable}. Using the method of Friend et al. \cite{Friend11,BrownThesis}, the linearity of the responses of the target PMTs were evaluated and shown to be linear with one exception. One of the PMTs coupled to the 2.54~cm EJ-232Q scintillator (Hamamatsu H1949-50 PMT biased at 2050~V) exhibited a nonlinear response and waveforms were corrected on a per sample basis. This PMT was several years older than the other H1949-50 PMTs, and the manufacturer confirmed that variation in linearity is to be expected across different PMTs of the same model \cite{PrivCommHamamatsu}.

\begin{table*} 
\centering
	\renewcommand{\arraystretch}{1.2}
	\begin{tabular}{ccccc}
	\hline
		\hline
		Material & Scatter ID &		x (cm)			&	y (cm)			&	z (cm)		\\ \hline
		EJ-230	& - 		&		57.7	$\pm$	0.5& $	0.0	\pm	0.5	$   &$0.0	\pm	0.5	$\\ 
		EJ-309	& 0		&	205.8	$\pm$	1	& $	80.2	\pm	1	$&$	-1.8	\pm	0.2	$\\ 
		EJ-309	& 1 		&	243.4	$\pm$	1	&$	69.5	\pm	1	$&$	18.4	\pm	0.2	$\\ 
		EJ-309	& 2 		&	149.9 $\pm$	    1	&$	98.7	\pm	1	$&$	20.0	\pm	0.2	$\\ 
		EJ-309	& 3 		&	171.9	$\pm$	1	&$	92.5	\pm	1	$&$	18.2	\pm	0.2	$\\ 
		EJ-309	& 4		&	290.0	$\pm$	1	&$	44.9	\pm	1	$&$ 16.9	\pm	0.2	$\\ 
		EJ-309	& 5		&	290.0 $\pm$	    1	&$	44.9	\pm	1	$&$	52.0	\pm	0.2	$\\ 
		EJ-309	& 6 		&	170.1 $\pm$	    1	&$	92.6	\pm	1	$&$	-2.6	\pm	0.2	$\\ 
		EJ-309	& 7		&	217.3 $\pm$	    1	&$	78.0	\pm	1	$&$	18.6	\pm	0.2	$\\ 
		EJ-309	& 8		&	253.4 $\pm$	    1	&$	65.5	\pm	1	$&$ -1.9	\pm	0.2	$\\ 
		EJ-309	& 9 		&	194.0 $\pm$	    1	&$	85.4	\pm	1	$&$	14.3	\pm	0.2	$\\ 
		EJ-309	& 10		&	272.4 $\pm$	    1	&$	59.5	\pm	1	$&$	18.5\pm	0.2	$\\ 
		\hline
		\hline
	\end{tabular}
	\caption{Detector locations and associated uncertainties for EJ-230. \label{locationsEJ230}}
\end{table*}

\begin{table*} 
\centering
	\renewcommand{\arraystretch}{1.2}
	\begin{tabular}{ccccc}
		\hline
		\hline
		Material & Scatter ID &		x (cm)			&	y (cm)			&	z (cm)		\\ \hline
		EJ-232 (5.08~cm h.)	& - 		&		40.8	$\pm$	0.5& $	0.0	\pm	0.5	$   &$0.0	\pm	0.5	$\\ 
		EJ-232 (2.54~cm h.)	& - 		&		40.6	$\pm$	0.5& $	0.0	\pm	0.5	$   &$0.0	\pm	0.5	$\\
		EJ-309	& 0		&	146.0	$\pm$	1	& $	106.3	\pm	1	$&$	10.6	\pm	0.2	$\\ 
		EJ-309	& 1 		&	154.1	$\pm$	1	&$	98.0	\pm	1	$&$	-11.4	\pm	0.2	$\\ 
		EJ-309	& 2 		&	165.3 $\pm$	    1	&$	87.4	\pm	1	$&$	10.6	\pm	0.2	$\\ 
		EJ-309	& 3 		&	173.4	$\pm$	1	&$	78.7	\pm	1	$&$	-11.6	\pm	0.2	$\\ 
		EJ-309	& 4		&	186.1	$\pm$	1	&$	67.5	\pm	1	$&$ 9.6	\pm	0.2	$\\ 
		EJ-309	& 5		&	195.7 $\pm$	    1	&$	58.7	\pm	1	$&$	-9.5	\pm	0.2	$\\ 
		EJ-309	& 6 		&	205.1 $\pm$	    1	&$	49.8	\pm	1	$&$	9.1	\pm	0.2	$\\ 
		EJ-309	& 7		&	223.3 $\pm$	    1	&$	31.7	\pm	1	$&$	-9.5	\pm	0.2	$\\ 
		EJ-309	& 8		&	224.7 $\pm$	    1	&$	30.6	\pm	1	$&$ 9.6	\pm	0.2	$\\ 
		EJ-309	& 9 		&	300 $\pm$	    1	&$	27.5	\pm	1	$&$	9.8	\pm	0.2	$\\ 
		EJ-309	& 10		&	300 $\pm$	    1	&$	27.5	\pm	1	$&$	-9.2\pm	0.2	$\\ 
		\hline
		\hline
	\end{tabular}
	\caption{Detector locations and associated uncertainties for EJ-232. \label{locationsEJ232}}
\end{table*}

\begin{table*} 
\centering
	\renewcommand{\arraystretch}{1.2}
	\begin{tabular}{ccccc}
		\hline
		\hline
		Material & Scatter ID &		x (cm)			&	y (cm)			&	z (cm)		\\ \hline
		EJ-232Q (5.08~cm h.)	& - 		&		41.6	$\pm$	0.5& $	0.0	\pm	0.5	$   &$0.0	\pm	0.5	$\\ 
		EJ-232Q (2.54~cm h.)	& - 		&		42.3	$\pm$	0.5& $	0.0	\pm	0.5	$   &$0.0	\pm	0.5	$\\
		EJ-309	& 0		&	146.0	$\pm$	1	& $	106.3	\pm	1	$&$	10.9	\pm	0.2	$\\ 
		EJ-309	& 1 		&	153.8	$\pm$	1	&$	97.7	\pm	1	$&$	-11.1	\pm	0.2	$\\ 
		EJ-309	& 2 		&	165.4 $\pm$	    1	&$	87.2	\pm	1	$&$	11.6	\pm	0.2	$\\ 
		EJ-309	& 3 		&	173.1	$\pm$	1	&$	78.3	\pm	1	$&$	-11.2	\pm	0.2	$\\ 
		EJ-309	& 4		&	186.3	$\pm$	1	&$	67.6	\pm	1	$&$ 9.7	\pm	0.2	$\\ 
		EJ-309	& 5		&	195.6 $\pm$	    1	&$	58.4	\pm	1	$&$	-9.2	\pm	0.2	$\\ 
		EJ-309	& 6 		&	205.0 $\pm$	    1	&$	50.0	\pm	1	$&$	9.6	\pm	0.2	$\\ 
		EJ-309	& 7		&	223.0 $\pm$	    1	&$	31.4	\pm	1	$&$	-9.0	\pm	0.2	$\\ 
		EJ-309	& 8		&	224.7 $\pm$	    1	&$	30.7	\pm	1	$&$ 9.7	\pm	0.2	$\\ 
		EJ-309	& 9 		&	315.9 $\pm$	    1	&$	27.7	\pm	1	$&$	18.3	\pm	0.2	$\\ 
		EJ-309	& 10		&	315.9 $\pm$	    1	&$	27.7	\pm	1	$&$	-7.1\pm	0.2	$\\ 
		\hline
		\hline
	\end{tabular}
	\caption{Detector locations and associated uncertainties for EJ-232Q. \label{locationsEJ232Q}}
\end{table*}

\begin{table*} 
\centering
	\renewcommand{\arraystretch}{1.2}
	\begin{tabular}{cccc}
		\hline
		\hline
		Material & Height (cm) & PMT Model &		Bias Voltage (V) \\ \hline
		EJ-230 & 5.08 & H1949-51 & -1700 \\
		EJ-230 & 5.08 & H1949-51 & -1685  \\
		EJ-232 & 5.08 & H1949-51 & -1700  \\
		EJ-232 & 5.08 & H1949-51 & -1725  \\
		EJ-232Q & 5.08 & H1949-51 & -1975  \\
		EJ-232Q	& 5.08 & H1949-50 & -1900 \\
		EJ-232 & 2.54 & H1949-51 & -1800 \\
		EJ-232 & 2.54 & H1949-50 & -1650 \\
		EJ-232Q & 2.54 & H1949-50 & -2050 \\
		EJ-232Q & 2.54 & H1949-50 & -1900 \\
		\hline
		\hline
	\end{tabular}
	\caption{Details on the Hamamatsu PMTs used for each scintillator material. The anode signals are directly coupled. \label{pmtTable}}
\end{table*}

\section{Proton Light Yield Data}
\label{LYapp}
The relative PLY data for EJ-230, EJ-232, and EJ-232Q obtained using integration lengths of 180, 180, and 140~ns, respectively, are summarized in Tables \ref{resultsTable230}, \ref{resultsTable232}, and \ref{resultsTable232Q}.

\begin{table*}
	\centering
	\renewcommand{\arraystretch}{1.2}
	\setlength{\tabcolsep}{8pt}
	\begin{tabular}{cc}
		\hline
		\hline
		Proton recoil  & Light relative to 477 keV electron  \\
		          energy [MeV] &  [dimensionless]\\
		\hline
		0.217$_{-0.017}^{+0.017}$ & 0.0369 $\pm$ 0.0017 \\
		0.254$_{-0.019}^{+0.021}$ & 0.0446 $\pm$ 0.0020 \\
		0.299$_{-0.024}^{+0.024}$ & 0.0560 $\pm$ 0.0023 \\ 
		0.350$_{-0.028}^{+0.028}$ & 0.0720 $\pm$ 0.0030 \\ 
		0.410$_{-0.033}^{+0.032}$ & 0.0921 $\pm$ 0.0035 \\ 
		0.478$_{-0.036}^{+0.039}$ & 0.1170 $\pm$ 0.0042 \\ 
		0.560$_{-0.042}^{+0.045}$ & 0.1481 $\pm$ 0.0051 \\ 
		0.655$_{-0.050}^{+0.052}$ & 0.1900 $\pm$ 0.0065 \\ 
		0.766$_{-0.059}^{+0.062}$ & 0.2451 $\pm$ 0.0080 \\ 
		0.896$_{-0.068}^{+0.071}$ & 0.3141 $\pm$ 0.0102 \\ 
		1.045$_{-0.078}^{+0.080}$ & 0.4041 $\pm$ 0.0134 \\ 
		1.216$_{-0.091}^{+0.096}$ & 0.5177 $\pm$ 0.0152 \\ 
		1.418$_{-0.107}^{+0.113}$ & 0.6610 $\pm$ 0.0186 \\ 
		1.655$_{-0.124}^{+0.128}$ & 0.8435 $\pm$ 0.0235 \\ 
		1.932$_{-0.150}^{+0.156}$ & 1.0873 $\pm$ 0.0292 \\ 
		2.250$_{-0.162}^{+0.182}$ & 1.3699 $\pm$ 0.0369 \\ 
		2.627$_{-0.196}^{+0.216}$ & 1.7676 $\pm$ 0.0478 \\ 
		3.077$_{-0.234}^{+0.255}$ & 2.2232 $\pm$ 0.0567 \\ 
		3.590$_{-0.258}^{+0.286}$ & 2.8142 $\pm$ 0.0796 \\ 
		4.168$_{-0.292}^{+0.364}$ & 3.4658 $\pm$ 0.0812 \\ 
		\hline
		\hline
	\end{tabular}
	\caption{Relative proton light yield data for the EJ-230 scintillator (5.08~cm h.\ by 5.08~cm dia.\ cylindrical cell). Proton recoil energy bin widths are provided, as well as the light output uncertainties. The covariance matrix is available upon request.}
	\label{resultsTable230}
\end{table*}

\begin{table*}
	\centering
	\renewcommand{\arraystretch}{1.2}
	\setlength{\tabcolsep}{8pt}
	\begin{tabular}{cc|cc}
		\hline
		\hline
		\multicolumn{2}{c|}{EJ-232, 5.08~cm h.} & \multicolumn{2}{c}{EJ-232, 2.54~cm h.}   \\
		Proton recoil   & Light relative to 477 keV electron  & Proton recoil  & Light relative to 477 keV electron  \\
		          energy [MeV] &  [dimensionless] & energy [MeV] & [dimensionless] \\
		\hline
		0.341$_{-0.035}^{+0.033}$ & 0.0620 $\pm$ 0.0029 &
0.609$_{-0.059}^{+0.051}$ & 0.1459 $\pm$ 0.0078 \\ 
		0.412$_{-0.038}^{+0.044}$ & 0.0818 $\pm$ 0.0037 &
0.727$_{-0.068}^{+0.058}$ & 0.1986 $\pm$ 0.0106 \\ 
		0.505$_{-0.049}^{+0.045}$ & 0.1141 $\pm$ 0.0054 &
0.856$_{-0.072}^{+0.070}$ & 0.2555 $\pm$ 0.0121 \\ 
		0.603$_{-0.054}^{+0.056}$ & 0.1493 $\pm$ 0.0055 & 
1.005$_{-0.079}^{+0.083}$ & 0.3320 $\pm$ 0.0149 \\
		0.718$_{-0.059}^{+0.067}$ & 0.1961 $\pm$ 0.0077 &
1.186$_{-0.097}^{+0.091}$ & 0.4340 $\pm$ 0.0204 \\ 
		0.859$_{-0.074}^{+0.067}$ & 0.2658 $\pm$ 0.0096 &
1.392$_{-0.114}^{+0.115}$ & 0.5698 $\pm$ 0.0240 \\ 
		1.005$_{-0.079}^{+0.084}$ & 0.3419 $\pm$ 0.0120 &
1.638$_{-0.132}^{+0.121}$ & 0.7480 $\pm$ 0.0310 \\ 
		1.182$_{-0.093}^{+0.096}$ & 0.4454 $\pm$ 0.0153 &
1.914$_{-0.154}^{+0.155}$ & 0.9607 $\pm$ 0.0402 \\ 
		1.389$_{-0.111}^{+0.118}$ & 0.5772 $\pm$ 0.0179 &
2.241$_{-0.171}^{+0.169}$ & 1.2420 $\pm$ 0.0480 \\ 
		1.629$_{-0.123}^{+0.130}$ & 0.7472 $\pm$ 0.0222 &
2.598$_{-0.189}^{+0.205}$ & 1.5696 $\pm$ 0.0625 \\
		1.908$_{-0.148}^{+0.161}$ & 0.9638 $\pm$ 0.0280 & 
3.018$_{-0.215}^{+0.240}$ & 1.9709 $\pm$ 0.0748 \\ 
		2.236$_{-0.167}^{+0.174}$ & 1.2549 $\pm$ 0.0345 &
3.503$_{-0.245}^{+0.288}$ & 2.4522 $\pm$ 0.0894  \\ 
		2.594$_{-0.185}^{+0.209}$ & 1.5885 $\pm$ 0.0418 &  &  \\
		3.017$_{-0.214}^{+0.241}$ & 2.0124 $\pm$ 0.0529 &  &  \\ 
		3.501$_{-0.243}^{+0.289}$ & 2.5198 $\pm$ 0.0612 &  &  \\ 
		4.064$_{-0.274}^{+0.337}$ & 3.1164 $\pm$ 0.0754 &  &  \\ 
		\hline
		\hline
	\end{tabular}
	\caption{Relative proton light yield data for the EJ-232 scintillators. Two different cylindrical scintillator cell sizes were evaluated, both 5.08~cm dia. Proton recoil energy bin widths are provided, as well as the light output uncertainties. Covariance matrices are available upon request.}
	\label{resultsTable232}
\end{table*}

\begin{table*}
	\centering
	\renewcommand{\arraystretch}{1.2}
	\setlength{\tabcolsep}{8pt}
	\begin{tabular}{cc|cc}
		\hline
		\hline
		\multicolumn{2}{c|}{EJ-232Q, 5.08~cm h.} & \multicolumn{2}{c}{EJ-232Q, 2.54~cm h.}   \\
		Proton recoil   & Light relative to 477 keV electron  & Proton recoil  & Light relative to 477 keV electron  \\
		          energy [MeV] &  [dimensionless] & energy [MeV] & [dimensionless] \\
		\hline
		0.413$_{-0.041}^{+0.039}$ & 0.0911 $\pm$ 0.0056 & 0.412$_{-0.040}^{+0.040}$ & 0.0912 $\pm$ 0.0047 \\ 
		0.500$_{-0.047}^{+0.045}$ & 0.1229 $\pm$ 0.0071 & 0.499$_{-0.046}^{+0.046}$ & 0.1235 $\pm$ 0.0061 \\ 
		0.598$_{-0.053}^{+0.054}$ & 0.1659 $\pm$ 0.0089 & 0.597$_{-0.052}^{+0.055}$ & 0.1640 $\pm$ 0.0073 \\ 
		0.714$_{-0.062}^{+0.061}$ & 0.2186 $\pm$ 0.0115 & 0.713$_{-0.060}^{+0.062}$ & 0.2187 $\pm$ 0.0098 \\ 
		0.841$_{-0.066}^{+0.071}$ & 0.2829 $\pm$ 0.0133 & 0.841$_{-0.067}^{+0.071}$ & 0.2847 $\pm$ 0.0118 \\ 
		0.991$_{-0.079}^{+0.082}$ & 0.3680 $\pm$ 0.0173 & 0.990$_{-0.078}^{+0.082}$ & 0.3701 $\pm$ 0.0155 \\ 
		1.165$_{-0.093}^{+0.094}$ & 0.4761 $\pm$ 0.0209 & 1.163$_{-0.091}^{+0.096}$ & 0.4773 $\pm$ 0.0173 \\ 
		1.366$_{-0.106}^{+0.112}$ & 0.6126 $\pm$ 0.0274 & 1.366$_{-0.106}^{+0.113}$ & 0.6134 $\pm$ 0.0214 \\ 
		1.602$_{-0.124}^{+0.130}$ & 0.7891 $\pm$ 0.0322 & 1.600$_{-0.122}^{+0.133}$ & 0.7857 $\pm$ 0.0241 \\ 
		1.876$_{-0.143}^{+0.150}$ & 1.0077 $\pm$ 0.0402 & 1.876$_{-0.143}^{+0.150}$ & 1.0115 $\pm$ 0.0291 \\ 
		2.193$_{-0.168}^{+0.173}$ & 1.2901 $\pm$ 0.0504 & 2.191$_{-0.166}^{+0.175}$ & 1.2835 $\pm$ 0.0360 \\ 
		2.546$_{-0.179}^{+0.204}$ & 1.6236 $\pm$ 0.0665 & 2.548$_{-0.181}^{+0.202}$ & 1.6327 $\pm$ 0.0427 \\ 
		2.960$_{-0.210}^{+0.240}$ & 2.0428 $\pm$ 0.0807 & 2.961$_{-0.211}^{+0.239}$ & 2.0389 $\pm$ 0.0528 \\ 
		3.436$_{-0.236}^{+0.277}$ & 2.5203 $\pm$ 0.0928 & 3.436$_{-0.235}^{+0.277}$ & 2.5331 $\pm$ 0.0650 \\ 
		3.974$_{-0.261}^{+0.324}$ & 3.0846 $\pm$ 0.1317 & 3.971$_{-0.258}^{+0.326}$ & 3.1103 $\pm$ 0.0764 \\ 
		4.580$_{-0.283}^{+0.390}$ & 3.7138 $\pm$ 0.1343 & 4.589$_{-0.292}^{+0.381}$ & 3.7681 $\pm$ 0.0861 \\ 
				
		\hline
		\hline
	\end{tabular}
	\caption{Relative proton light yield data for the EJ-232Q scintillators. Two different cylindrical scintillator cell sizes were evaluated, both 5.08~cm dia. Proton recoil energy bin widths are provided, as well as the light output uncertainties. Covariance matrices are available upon request.}
	\label{resultsTable232Q}
\end{table*}

\section*{Acknowledgments}
The authors would like to thank the 88-Inch Cyclotron operations and facilities staff for their help in performing these experiments. 

\bibliographystyle{IEEEtran}
\bibliography{IEEEabrv,./EJ23xArXiv.bib}

\end{document}